\documentclass{nature}
\usepackage{graphicx,psfrag}
\usepackage{mathrsfs}
\usepackage{amsmath,amsfonts,amssymb}
\usepackage{multirow}
\usepackage{comment}
\usepackage{units}
\usepackage{enumerate}
\usepackage[normalem]{ulem} 
\usepackage{longtable}
\usepackage[rgb]{xcolor}
\usepackage{gensymb}
\usepackage{adjustbox}
\usepackage[none]{hyphenat}
\usepackage{multibib}
\newcites{New}{References}

\title{General Relativistic Orbital Decay in a 7 Minute Orbital Period Eclipsing Binary System}

\author{Kevin B. Burdge$^{1}$$^{*}$, Michael W. Coughlin$^{1}$, Jim Fuller$^{1}$, Thomas Kupfer$^{2}$, Eric C. Bellm$^{3}$, Lars Bildsten$^{2,4}$, Matthew J. Graham$^{1}$, David L. Kaplan$^{5}$, Jan van Roestel$^{1}$, Richard G. Dekany$^{6}$, Dmitry A. Duev$^{1}$, Michael Feeney$^{6}$, Matteo Giomi$^{7}$, George Helou$^{8}$, Stephen Kaye$^{6}$, Russ R. Laher$^{8}$, Ashish A. Mahabal$^{1}$, Frank J. Masci$^{8}$, Reed Riddle$^{6}$, David L. Shupe$^{8}$, Maayane T. Soumagnac$^{9}$, Roger M. Smith$^{6}$, Paula Szkody$^{3}$, Richard Walters$^{6}$ \& S. R. Kulkarni$^1$, Thomas A. Prince$^1$
}

\begin{document}

\maketitle

\begin{affiliations}
 \item Division of Physics, Mathematics and Astronomy, California Institute of Technology, Pasadena, CA, USA
 \item Kavli Institute for Theoretical Physics, University of California Santa-Barbara, Santa Barbara, CA, USA
 \item Department of Astronomy, University of Washington, Seattle, WA, USA
 \item Department of Physics, University of California, Santa Barbara, CA, USA
 \item Department of Physics, University of Wisconsin-Milwaukee, Milwaukee, WI, USA
 \item Caltech Optical Observatories, California Institute of Technology, Pasadena, CA, USA
 \item Humboldt-Universit\"{a}t zu Berlin, Berlin, Germany
 \item IPAC, California Institute of Technology, Pasadena, CA, USA
 \item Benoziyo Center for Astrophysics, Weizmann Institute of Science, Rehovot, Israel
\end{affiliations}

\begin{abstract}
General relativity\cite{1916SPAW.......688E} predicts that short orbital period binaries emit significant gravitational radiation, and the upcoming Laser Interferometer Space Antenna (LISA)\cite{amaro2017laser} is expected to detect tens of thousands of such systems\cite{nissanke2012gravitational}; however, few have been identified\cite{kupfer2018lisa}, and only one is eclipsing--the double white dwarf binary SDSS J065133.338+284423.37\cite{brown201112}, which has an orbital period of 12.75 minutes. Here, we report the discovery of an eclipsing double white dwarf binary system with an orbital period of only 6.91 minutes, ZTF J153932.16+502738.8. This system has an orbital period close to half that of SDSS J065133.338+284423.37, and an orbit so compact that the entire binary could fit within the diameter of the planet Saturn. The system exhibits a deep eclipse, and a double-lined spectroscopic nature. We observe rapid orbital decay, consistent with that expected from general relativity. ZTF J153932.16+502738.8 is a significant source of gravitational radiation close to the peak of LISA's sensitivity\cite{amaro2017laser}, and should be detected within the first week of LISA observations.
\end{abstract}

The Zwicky Transient Facility (ZTF)\cite{bellm2018zwicky}$^{,}$\cite{graham2019zwicky} is a northern-sky synoptic survey using the 48-inch Samuel Oschin Telescope at Palomar Observatory. In June 2018, we undertook an initial search for periodic sources among all of the 20 million ZTF lightcurves available at that time. The analysis identified ZTF J153932.16+502738.8 (henceforth referred to as ZTF J1539+5027), as a candidate binary system with a short orbital period. On the same night as identifying the candidate, an observation with the Kitt Peak 84-Inch Electron Multiplying Demonstrator (KPED)\cite{KPED} confirmed the discovery, and revealed a remarkably deep eclipse occurring precisely every $6.91$ minutes. Next, we used the high-speed imaging photometer CHIMERA\cite{harding2016chimera} on the 200-inch Hale telescope at Palomar Observatory to observe the system (Figure 1), confirming the deep primary eclipse, and revealing a shallow secondary eclipse.

\begin{figure}
\includegraphics[width=6.5in]{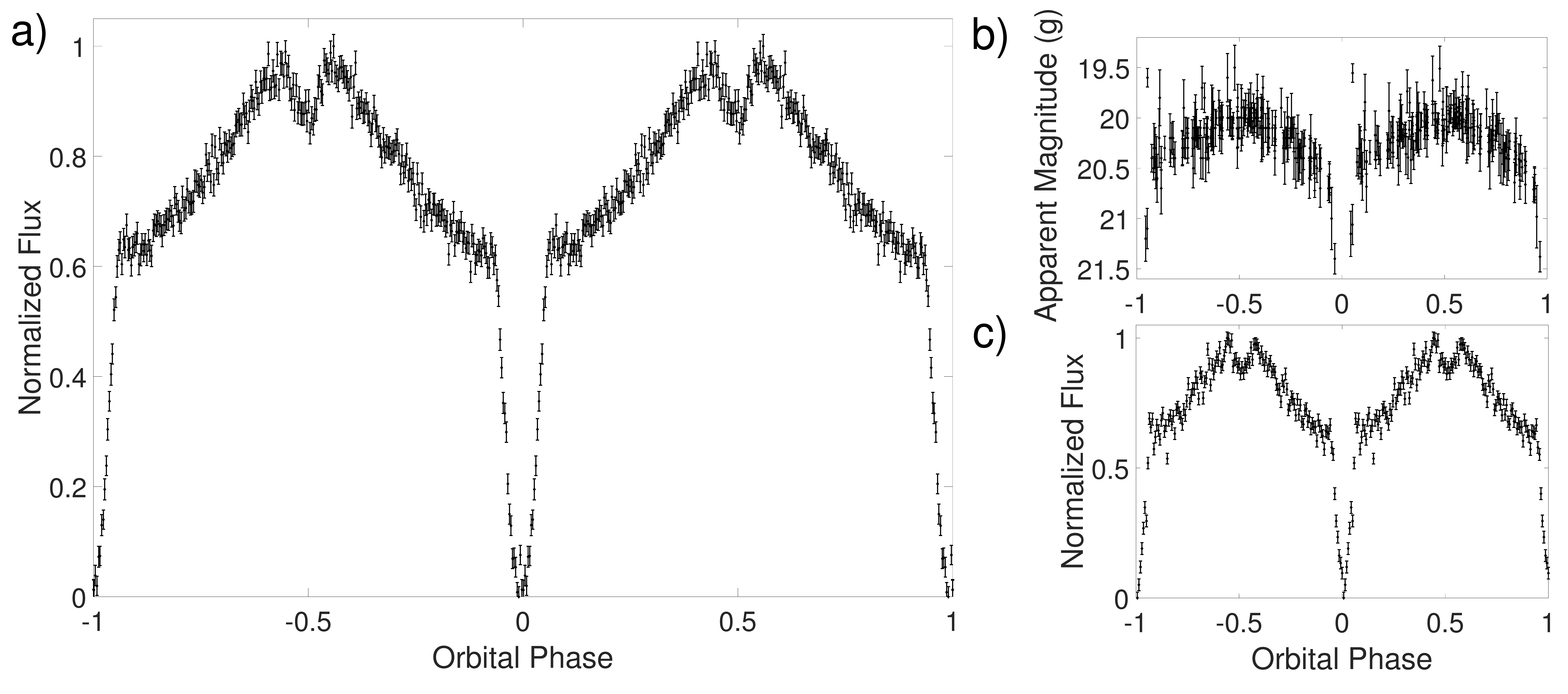}
\linespread{1.3}\selectfont{}
\caption{\textbf{Lightcurve of ZTF J1539+5027} \textbf{a)} The binned CHIMERA $g^\prime$ lightcurve of ZTF J1539+5027, phase-folded on the $6.91$ minute orbital period. At phase $0$, the lightcurve exhibits a deep primary eclipse, indicating that the hot primary star is producing most of the observed light. Outside of eclipse, there is a quasi-sinusoidal modulation because the primary star heavily irradiates one side of its companion. At phases $\pm0.5$, the secondary eclipse occurs as the hot primary transits the irradiated face of its companion. \textbf{b)} The phase-folded ZTF $g$-band lightcurve of the object. We were able to discover the object because of its periodic behavior. \textbf{c)} A binned $g^\prime$ lightcurve obtained with KPED, phase-folded on the orbital period. Error bars are $1\sigma$ intervals.}
\label{fig:chimera}
\end{figure}

The short orbital period means that the two components must be dense objects--white dwarfs. Because the primary eclipse is significantly deeper than the secondary eclipse, we can infer that one white dwarf (the primary) is hotter and more luminous than its companion (the secondary), as the detected flux is almost completely attenuated when the cooler object occults the hotter. By modelling the lightcurve (Methods), we can estimate the orbital inclination, $i$, the radius of the primary, $R_1$, and the secondary, $R_2$, relative to the semi-major axis of the orbit, $a$ (Methods).

Because of ZTF J1539+5027's extremely short orbital period, general relativity predicts that it will undergo rapid orbital decay due to the emission of gravitational radiation\cite{taylor1979measurements}. With CHIMERA and KPED, we can precisely measure the time of eclipse, and use these eclipse times to measure a changing orbital period. If a system has a constant orbital period derivative, we expect the deviation of eclipse times, $\Delta t_{eclipse}$, (compared to those of a system with constant orbital period) to grow quadratically in time. Equation \ref{eq:OC} \begin{equation} \label{eq:OC}
\Delta t_{eclipse}(t-t_0)= \Big(\frac{1}{2}\dot{f}(t_0)(t-t_0)^2+\frac{1}{6}\ddot{f}(t_0)(t-t_0)^3+... \Big) P(t_0)
\end{equation} illustrates this, where $t_0$ is the reference epoch, $P(t_0)$ is the orbital period at the reference epoch, $f(t_0), \dot{f}(t_0),$ $ \rm etc,$ are the orbital frequency and its time derivatives at the reference epoch, and $t-t_0$ is the time since the reference epoch.

We also used IRSA/IPAC\cite{masci2016ipac} to retrieve photometry from archival Palomar Transient Factory (PTF/iPTF) data\cite{law2009palomar} spanning 2009, 2010, 2011, and 2016. Figure 2 shows a fit of all of the timing epochs with a second order polynomial, which resulted in a highly significant detection of the orbital decay, corresponding to an orbital period derivative of $\dot{P}=(-2.373\pm0.005)\times10^{-11} \mathrm{s\,s}^{-1}$ (Table 1). The corresponding characteristic orbital decay timescale is: $\tau_c=\frac{3}{8}\frac{P}{|\dot{P}|}\approx 210,000$\,years.

\begin{figure}
\includegraphics[width=6.5in]{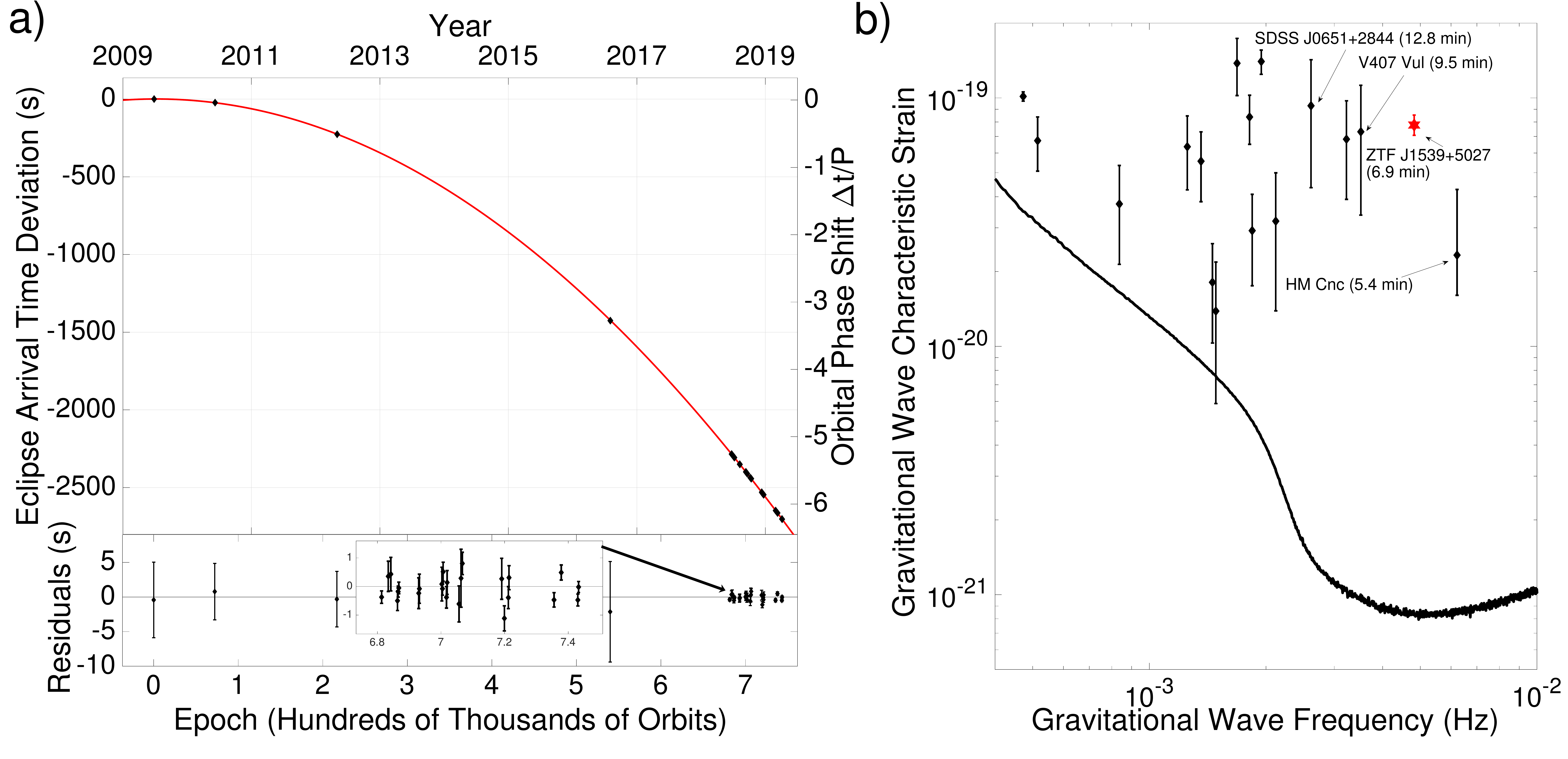}
\linespread{1.3}\selectfont{}
\caption{\textbf{Orbital decay and gravitational wave strain of ZTF J1539+5027} \textbf{a)} A 2nd order polynomial fit to the deviation of the measured eclipse times as a function of time, compared to a system with constant orbital period. The consistency with a quadratic deviation demonstrates that the orbital period decreases with time. The orbital decay inferred is consistent with that expected from gravitational wave emission. The initial four timing epochs come from PTF/iPTF photometry, and the remainder were obtained with CHIMERA and KPED. \textbf{b)} The characteristic gravitational wave strain and frequency for ZTF J1539+5027 (red star in the plot). See Table 1 for masses and the distance. The black diamonds are other known LISA sources, all of which are compact binaries\cite{kupfer2018lisa}. The smooth black curve is the expected sensitivity threshold of LISA after $4$ years of integration\cite{amaro2017laser}. For HM Cancri (right-most point) we have assumed a uniform prior in distance from $4.2$-$20$ kpc\cite{2010ApJ...711L.138R}$^,$\cite{bildsten2006thermal}. Error bars on panel \textbf{a)} are $1 \sigma$. Errors on panel \textbf{b)} are taken from\cite{kupfer2018lisa} for all points except ZTF J1539 and HM Cnc, which are $68\%$ CIs.}

\label{fig:pdot}
\end{figure}

To measure the orbital velocities of the white dwarfs in the binary, we obtained phase-resolved spectroscopy using the Low Resolution Imaging Spectrometer (LRIS)\cite{mccarthy1998blue} on the 10-m W.\ M.\ Keck I Telescope on Mauna Kea. These observations (Figure 3) revealed broad and shallow hydrogen absorption lines characteristic of a hot hydrogen-rich (DA) white dwarf associated with the bright primary, and within these absorption lines, narrower hydrogen emission lines apparently arising from the cooler secondary. The emission lines move out of phase with the absorption lines, making this a double-lined spectroscopic binary. There are also weak neutral helium absorption and emission lines that exhibit similar behavior. The Doppler shifts of the emission lines in the spectra track the cool secondary, suggesting that the emission lines are not associated with accretion onto the hot and compact primary, but instead arise from the irradiated surface of the secondary.

\begin{figure}
\includegraphics[width=6.5in]{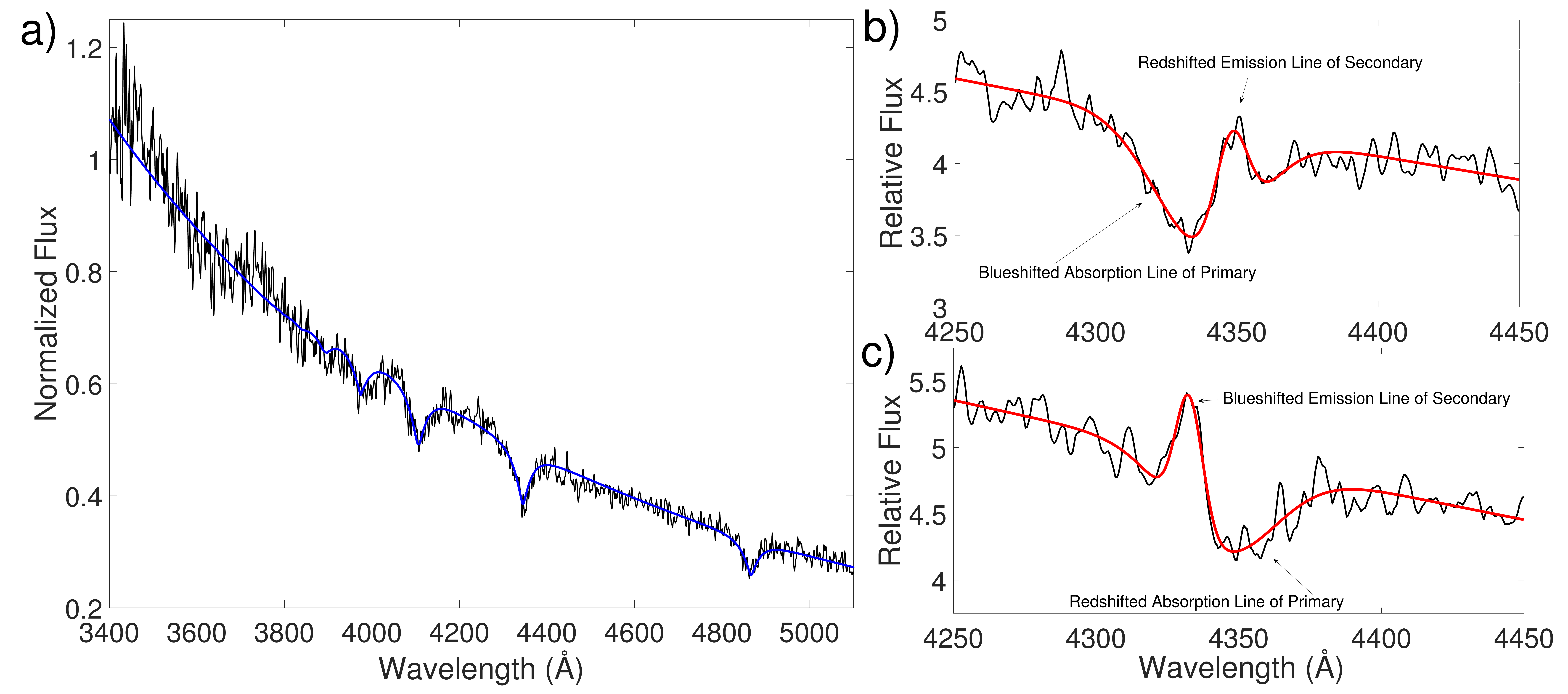}
\linespread{1.3}\selectfont{}
\caption{\textbf{Optical spectrum of ZTF J1539+5027} \textbf{a)} The LRIS spectrum of ZTF J1539+5027 immediately after the primary eclipse. This is an ideal phase to isolate the photosphere of the hot primary, because it minimizes flux contributed by the irradiated face of the secondary. The smooth blue line is a fit of a white dwarf atmospheric model to this spectrum, yielding an effective photospheric temperature of $T_{\rm eff,1}=48,900\pm900\,K$ and a logarithm of surface gravity $\log (g)_1=7.75\pm0.06\,\log(\rm{cm}\,\rm{s}^{-2})$ for the hot primary. \textbf{b)} and \textbf{c)} Two phase resolved spectra of the hydrogen $n=5$ to $n=2$ transition at $\approx4340\,\rm{\AA}$. The smooth red line is a double gaussian fit to the absorption and emission line used to measure the Doppler shifts of these features \textbf{b)} illustrates a phase in which the emission line associated with the cooler secondary is redshifted, while the absorption line associated with the primary is blueshifted. \textbf{c)} exhibits the opposing phase.
}
\label{fig:spectra}
\end{figure}

Using the spectroscopic observations, lightcurve modelling, and the orbital decay, we can constrain the masses of the white dwarfs in several ways: \textbf{(1):} With a mass-radius relation for the hot primary and constraints from lightcurve modelling, although this depends on parameters of white dwarf models, and only weakly constrains the mass of the secondary; \textbf{(2):} Using the spectroscopically measured radial velocity semi-amplitudes; however, this is challenging due to the blended absorption/emission lines, the latter depending on modelling irradiation effects and a substantial center of light correction; \textbf{(3):} With the chirp mass inferred from the measured orbital decay; however, this approach must account for potential tidal contributions. 

Because each of these methods rely on different model dependent assumptions, we chose to estimate physical parameters by combining these constraints (Figure 4). We present the physical parameters resulting from this analysis in Table 1, alongside those of SDSS J0651+2844\cite{brown201112,hermes2012rapid,kupfer2018lisa}. We conclude that the hot primary is a $\approx \! 0.6 \, M_\odot$ (likely carbon-oxygen) white dwarf, while the cool secondary is a $\approx \! 0.2 \, M_\odot$ (likely helium-core) white dwarf. 

\begin{figure}
    \centering
    \includegraphics[width=6.5in]{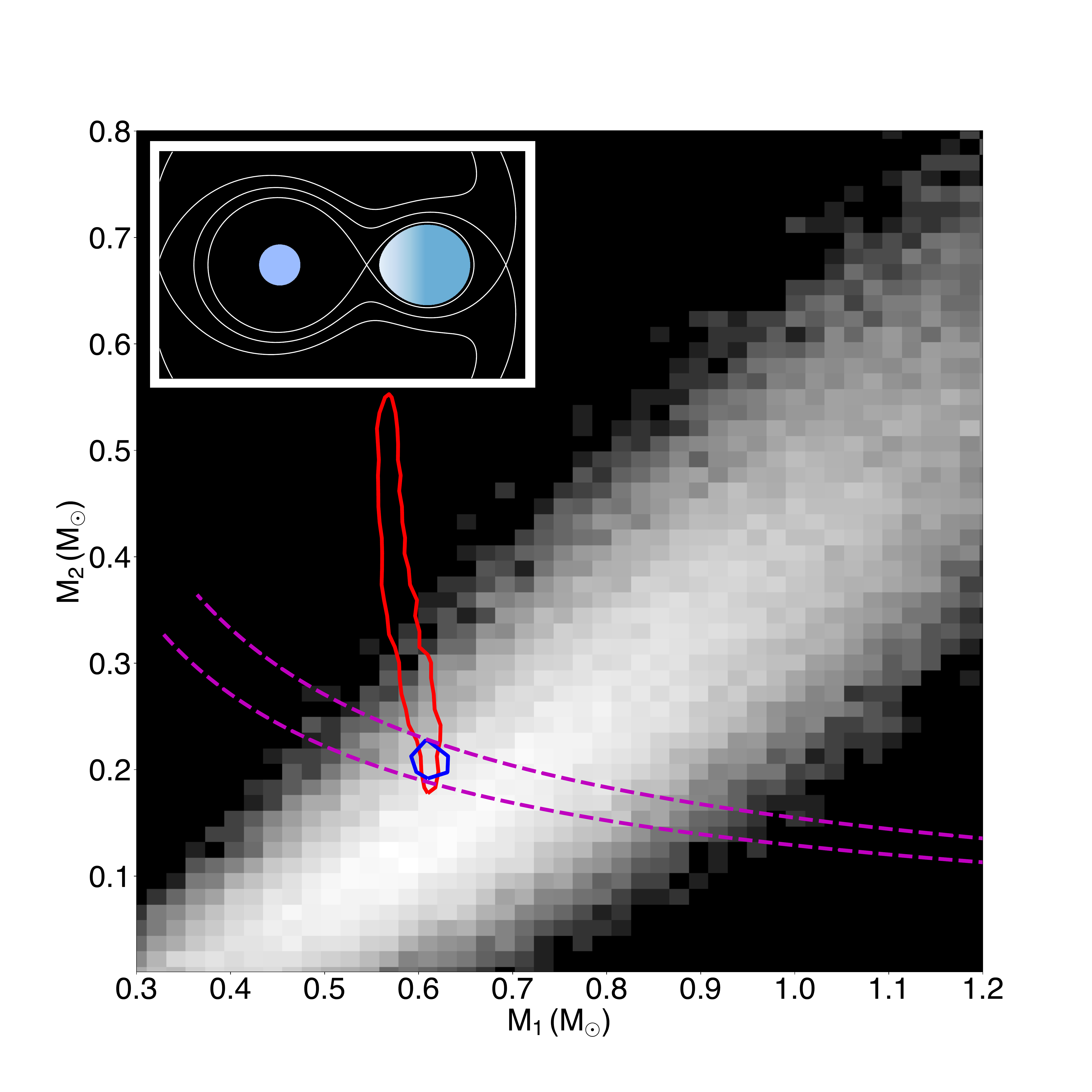}
    \linespread{1.3}\selectfont{}
    \caption{\textbf{Constraints on component masses in ZTFJ1539+5027} A graphical representation of the analysis used to derive final values for the masses of the two white dwarfs, with an inset illustrating the two components of the binary drawn to scale based on these parameters together with Roche potentials. The background white bins represent the constraint imposed by the measured spectroscopic radial velocity semi-amplitudes. The red line is a 50\% contour level of the constraint inferred from applying a mass-radius relation to the hot primary white dwarf\cite{AlCa2015}, and combining it with the ratio of the primary's radius to the semi-major axis, $\frac{R_1}{a}$, inferred from lightcurve modelling. The magenta dashed lines are constraints imposed by the measured chirp mass; the upper dashed line assumes orbital decay purely due to general relativity, whereas the lower dashed line includes a 10 percent tidal contribution. The blue line is a 50\% contour level representing a combination of all of these constraints (Table 1).}
    \label{fig:combined}
\end{figure}

Because of their remarkably short orbital period, the white dwarfs experience significant tidal distortion. Tidal energy dissipation may heat and spin up the white dwarfs, in addition to increasing the orbital decay rate. Based on theoretical predictions\cite{fuller2013dynamical}, tidal torques likely cause the spin periods of the white dwarfs to synchronize with the orbital period. We expect that tidal energy dissipation may increase the orbital decay rate by $\approx \!7 \%$ relative to gravitational wave emission alone (Methods), though we cannot currently measure this due to uncertainty in the white dwarf masses. Future detection of the second derivative of the orbital decay rate, $\ddot{f}$, will enable a direct measurement of the tidal contribution to the orbital decay.

When LISA\cite{amaro2017laser} first begins to operate (circa 2034), we estimate that it will detect ZTF J1539+5027 within a week. At the end of LISA's 4-year mission, we estimate the SNR will be $143^{+13}_{-13}$ (Methods). Not only does the source radiate high-strain gravitational waves, but LISA's sensitivity peaks at about $5$\,mHz, close to the $4.8$\,mHz gravitational wave frequency of this source, resulting in an exceptionally large SNR.
This system will serve as a crucial ``verification source'' for LISA\cite{amaro2017laser}, because its well-constrained inclination predicts the relative amplitude of the signal in the two gravitational wave polarizations\cite{shah:13}, and the precisely measured orbital decay already tightly constrains its chirp mass.

ZTF J1539+5027 poses challenges for models of binary evolution and the physics of accretion. The spectroscopically measured temperature of the hot primary is $T_{\rm eff,1} = 48,900 \pm 900\, {\rm K}$. The cooling age of such a hot white dwarf is $\approx 2.5$ million years\cite{holberg2006calibration}, significantly shorter than the $>200$ million year cooling age of the secondary\cite{2016A&A...595A..35I}. Thus, some recent heating must have occurred. Tidal heating could increase the surface temperature of the primary to nearly $50,000 \, {\rm K}$, though more realistic calculations (Methods) suggest temperatures closer to half this value. A more plausible explanation is that the heating is due to recent accretion, especially since the radius of the secondary indicates it is on the brink of Roche lobe overflow. Such accretion could heat the primary to its observed temperature for accretion rates of $\dot{M} \gtrsim 10^{-9} M_\odot \, {\rm yr}^{-1}$ (Methods).

However, we see no evidence for active accretion. The only other known binary systems with orbital periods shorter than 10 minutes, V407 Vul ($P \approx 9.5$ minutes)\cite{ramsay2000detection} and HM Cancri ($P \approx 5.4$ minutes)\cite{2010ApJ...711L.138R}, were discovered because of periodic X-ray emission, thought to arise from a hot spot formed by direct impact accretion\cite{marsh2002v407}. Unlike the other two sub-10-minute binaries, ZTF J1539+5027 exhibits no detectable X-ray flux. Based on an upper limits from observations by the XRT X-ray telescope on the Neil Gehrels Swift Observatory\cite{gehrels2004swift}, the EPIC-pn instrument on the XMM Newton Observatory\cite{jansen2001xmm}, and optical constraints, we have estimated an upper limit of $\dot{M} < 2\times10^{-8} M_\odot \,{\rm yr}^{-1}$, contributing $10\%$ of its energy to a hot spot. Reaching this upper limit requires fine-tuning the accretion hot spot temperature (Methods). Active accretion could still be occurring if the accretion energy is channeled primarily into heating the optically thick atmosphere of the accretor, and is then radiated at the $50,000\, \rm K$ observed.
 
It is also possible that the accretion proceeds intermittently. One way of temporarily halting accretion could be a recent nova eruption on the surface of the primary. We expect that a $0.6 \, M_\odot$ white dwarf accreting from a companion at a rate of $\dot{M} \sim 10^{-9} M_\odot \, {\rm yr}^{-1}$ should experience\cite{wolf:13} recurrent novae on timescales of $\sim 10^{5} \, {\rm yr}$. However, while mass transfer may temporarily cease after each nova, calculations suggest that it is unlikely to catch the system in this short-lived phase (Methods). We conclude that the most likely scenarios are intermittent accretion with a mechanism allowing for phases of little-to-no mass transfer, or active accretion in which the accretion energy is radiated almost entirely in the ultraviolet and optical.

The orbit of ZTF J1539+5027 will continue to decay for $\approx130,000$ years until it reaches a period of $\approx$5 minutes at which point the degenerate core of the secondary will begin to expand in response to mass loss, dramatically increasing the rate of mass transfer\cite{kaplan:12}. If the mass transfer is stable, which is likely based on the mass ratio\cite{marsh:04} of $q \sim 1/3$, the binary will evolve into an AM CVn system and the orbital period will increase. Alternatively, unstable mass transfer would result in a merger that could produce an R Cor Bor star\cite{paczynski1971evolution}, or, less likely, a detonation of accreted helium on the primary could lead to a double-detonation that disrupts the primary\cite{shen2018sub}. 
\

\

\begin{table}
\centering
\caption{Table of Parameters}
\medskip
\begin{tabular}{ccc}
           &  \bf{ZTF J1539+5027} & \bf{SDSS J0651+2844} \\
$M_1$&  $0.610^{+0.017}_{-0.022}$\,\(M_\odot\)&  $0.49^{+0.02}_{-0.02}$\,\(M_\odot\)\\
$M_2$ & $0.210^{+0.014}_{-0.015}$\,\(M_\odot\)& $0.247^{+0.015}_{-0.015}$\,\(M_\odot\)\\
$R_1$& $1.562^{+0.038}_{-0.038}\times10^{-2}$\,\(R_\odot\)& $1.42^{+0.1}_{-0.1}\times10^{-2}$\,\(R_\odot\)\\
$R_2$& $3.140^{+0.054}_{-0.052}\times10^{-2}$ \(R_\odot\)& $3.71^{+0.12}_{-0.12}\times10^{-2}$ \(R_\odot\) \\
$a$& $11.218^{+0.080}_{-0.082}\times10^{-2}$ \(R_\odot\)& $16.48^{+0.39}_{-0.43}\times10^{-2}$ \(R_\odot\) \\
$i$& $84.15^{+0.64}_{-0.57}\, \rm degrees $& $86.9^{+1.6}_{-1.0}\, \rm degrees $ \\
$T_0$&  $2458305.6827886\pm0.0000012\,\rm BJD_{TDB}$&  $2455652.5980910\pm0.0000084\,\rm BJD_{TDB}$ \\
$P$ & $414.7915404\pm0.0000029\,\rm s$& $765.206543\pm0.000055\,\rm s$  \\
$\dot{P}$& $(-2.373\pm0.005)\times10^{-11}\,\rm s\,\rm s^{-1}$& $(-0.98\pm0.28)\times10^{-11}\,\rm s\,\rm s^{-1}$ \\
$d$ & $2.34\pm0.14\, \rm kpc \,(spec.)$& $1.0\pm0.1\, \rm kpc\,(spec.)$ $0.9\pm0.5\, \rm kpc\,(parallax)$  \\
$T_{\rm eff,1}$ & $48,900\pm900 \, {\rm K}$ & $8,700\pm500 \, {\rm K}$ \\
$T_{\rm eff,2}$ & $<10,000 \, {\rm K}$ & $16,530\pm200 \, {\rm K}$  \\
$\log (g)_1$ & $7.75\pm0.06\,\log(\rm cm\,\rm s^{-2})$ &   \\
$\log (g)_2$ &  & $6.76\pm0.04\,\log(\rm cm\,\rm s^{-2})$  \\
$K_1$ & $292^{+254}_{-283}\,\rm km\,\rm s^{-1}$ &   \\
$K_2$ & $961^{+178}_{-139}\,\rm km\,\rm s^{-1}$ & $616.9^{+5.0}_{-5.0}\,\rm km\,\rm s^{-1}$  \\
$4\, \rm Yr\, LISA\, SNR$ & $143^{+14}_{-13}$ & $94^{+12}_{-10}\,\rm (using\, spec.\, distance)$  \\
$\rm References$ &  & \cite{kupfer2018lisa,brown201112,hermes2012rapid}  \\
\end{tabular}
\end{table}

\

\

\newpage

\begin{methods}

\section{Summary of Observations}

Extended Data Table 1 provides a summary of all observations used in our analysis.

\section{Period Finding}
We identified ZTF J1539+5027 by using the conditional entropy\cite{graham2013using} period finding algorithm on 20 million available ZTF lightcurves containing more than 50 epochs as of June 5, 2018, which originated exclusively from the ZTF collaboration's extragalactic high cadence survey (Bellm et al., PASP in press). These lightcurves correspond to approximately 10 million sources, each with a ZTF-$g$ and ZTF-$r$ lightcurve.

ZTF J1539+5027 exhibited the most significant signal of all the objects whose strongest (lowest entropy) period fell in the 6--7 minute range. We used a Graphics Processing Unit (GPU) implementation of the conditional entropy algorithm included in the \texttt{pycuda} based \texttt{cuvarbase} package (available on GitHub). The algorithm was executed on a pair of NVIDIA GTX 1080 Ti GPUs.

\section{Lightcurve Modelling}

To model the lightcurve, we used data from three nights of CHIMERA $g^\prime$ observations (July 5-7 2018), with a total of $12,999$ individual $3$-s exposures.

To fit the CHIMERA data, we used the \texttt{ellc} package\cite{maxted2016ellc} to model the lightcurve and fit for the ratio of the radii to the semi-major axis, $r_{1}=R_{1}/a$, $r_{2}=R_{2}/a$, inclination, $i$, mass ratio, $q=\frac{M_2}{M_1}$, surface brightness ratio, $J$, of the unheated face of the secondary compared to the hot primary, and the mid-eclipse time of the primary eclipse, $t_0$. We adopted a linear limb darkening model, with limb darkening coefficients for the primary ($\rm ldc_1$) and secondary ($\rm ldc_2$). We treated the primary as a spherical object, but invoked a Roche approximation for the secondary. We also included a gravity darkening coefficient for the secondary ($\rm gdc_2$) and a single free heating parameter ($\rm heat_2$) to attempt to fit the sinusoidal variation due to the irradiation of the secondary, which acts as an albedo of the secondary, but in this system must be larger than $1$ to achieve a good fit because a significant amount of incident ultraviolet light is reprocessed into optical wavelengths. 

We allowed the limb and gravity darkening coefficients to vary as free parameters in the fit, using uniform priors for each with values of $\rm ldc_1=0.15\pm{0.15}$, $\rm ldc_2=0.4\pm{0.2}$, and  $\rm gdc_2=0.6\pm{0.1}$, based on extrapolations of existing models\cite{gianninas2013limb}$^{,}$\cite{2011A&A...529A..75C}.

We performed the final fit using the period derived from the quadratic fit to the timing epochs (Table 1). We left $r_{1}$, $r_{2}$, $i$, $J$, $q$, $t_{0}$, $\rm ldc_1$, $\rm ldc_2$, $\rm gdc_2$, and $\rm heat_2$ as free parameters. Extended Data Figure 1 illustrates the corner plots from this fit, but excludes $t_{0}$, which did not exhibit significant covariance with any other parameter. We fixed the eccentricity to $0$, because measuring this quantity depends on the shape of the secondary eclipse, which in turn depends on the poorly understood irradiation of the secondary. We ruled out the possibility of significant eccentricity, because we failed to detect any sign of apsidal precession in the eclipse time measurements, and furthermore, we tried fitting for this using the lightcurve modelling, and found a value consistent with $0$. Because we lack a good physical model for the irradiation of the secondary, we also do not account for Doppler beaming, which we expect to alter the shape of the lightcurve at the few percent level, peaking at phases between the eclipses, where the irradiation dominates the behavior of the lightcurve.

\begin{figure}
\includegraphics[width=6.4in]{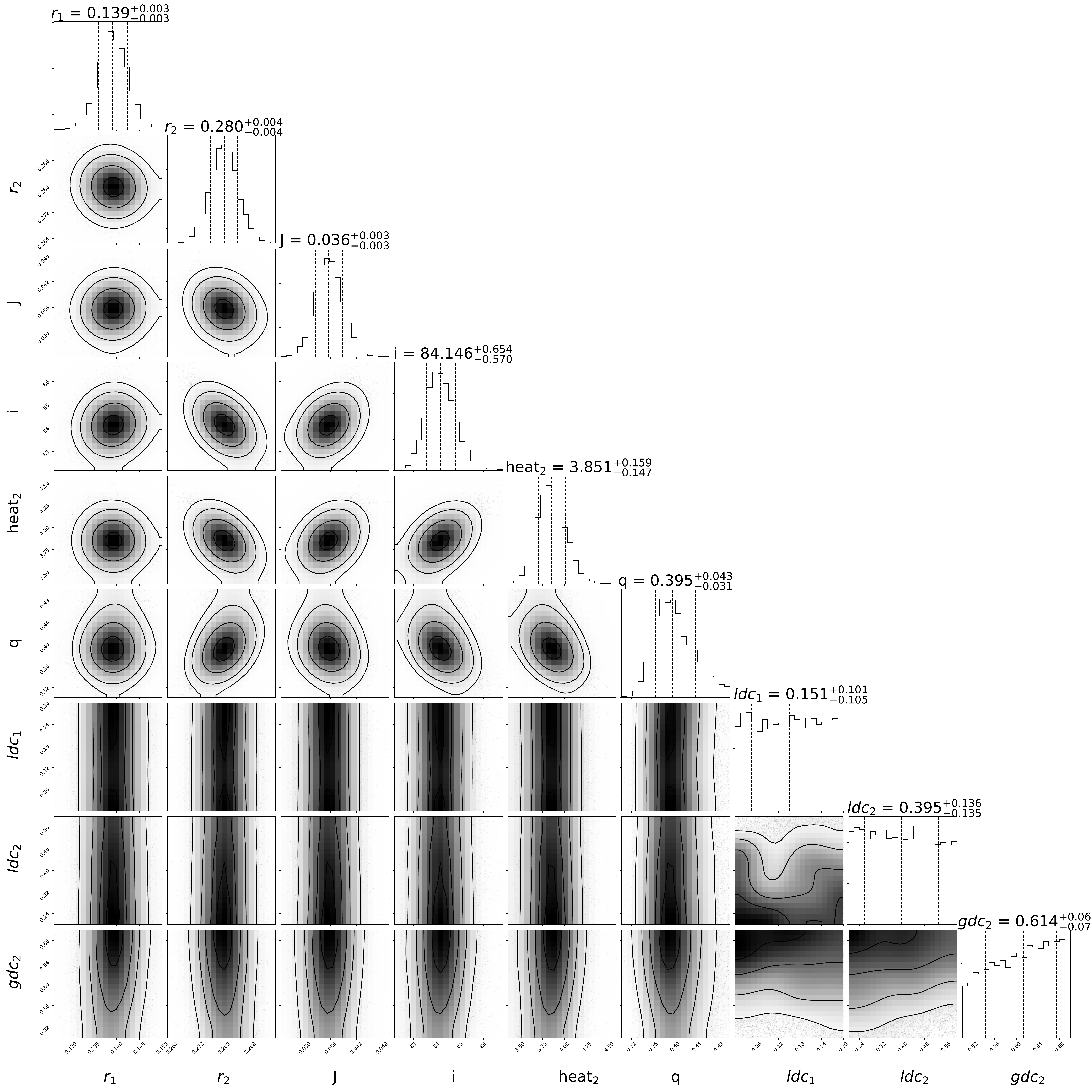}
\linespread{1.3}\selectfont{}
\caption{\textbf{Extended Data Figure 1: Corner plots of lightcurve modelling} The corner plots of the lightcurve fit to $12,999$ $g^\prime$ epochs taken with CHIMERA on July 5, 6, and 7 2018. Please note that the two limb darkening coefficients, as well as gravity darkening of the secondary (bottom three panels), were allowed to vary to ensure that assumptions regarding them were not strongly covariant with the other physical quantities of interest.}
\label{fig:LCCorner}
\end{figure}

\section{Orbital Decay}

To measure the orbital decay, we independently fit each night of KPED and CHIMERA data for mid-eclipse times using the lightcurve modelling performed with \texttt{ellc}. We convert all timestamps to Barycentric Julian Dates (BJD) in Barycentric Dynamical Time (TDB) to achieve the required timing precision. Because the eclipse time is not strongly covariant with model dependent parameters such as gravity and limb darkening, we omitted these from the fits to reduce complexity (but still fitting for all other parameters described in the previous section). We also extracted photometry from archival PTF/iPTF data, and because of significant smearing due to the exposure time of $60$-s (particularly for the primary eclipse), we extracted timing epochs by performing a least squares fit of a sinusoid to this data (Extended Data Figure 2).

\begin{figure}
\includegraphics[width=6.4in]{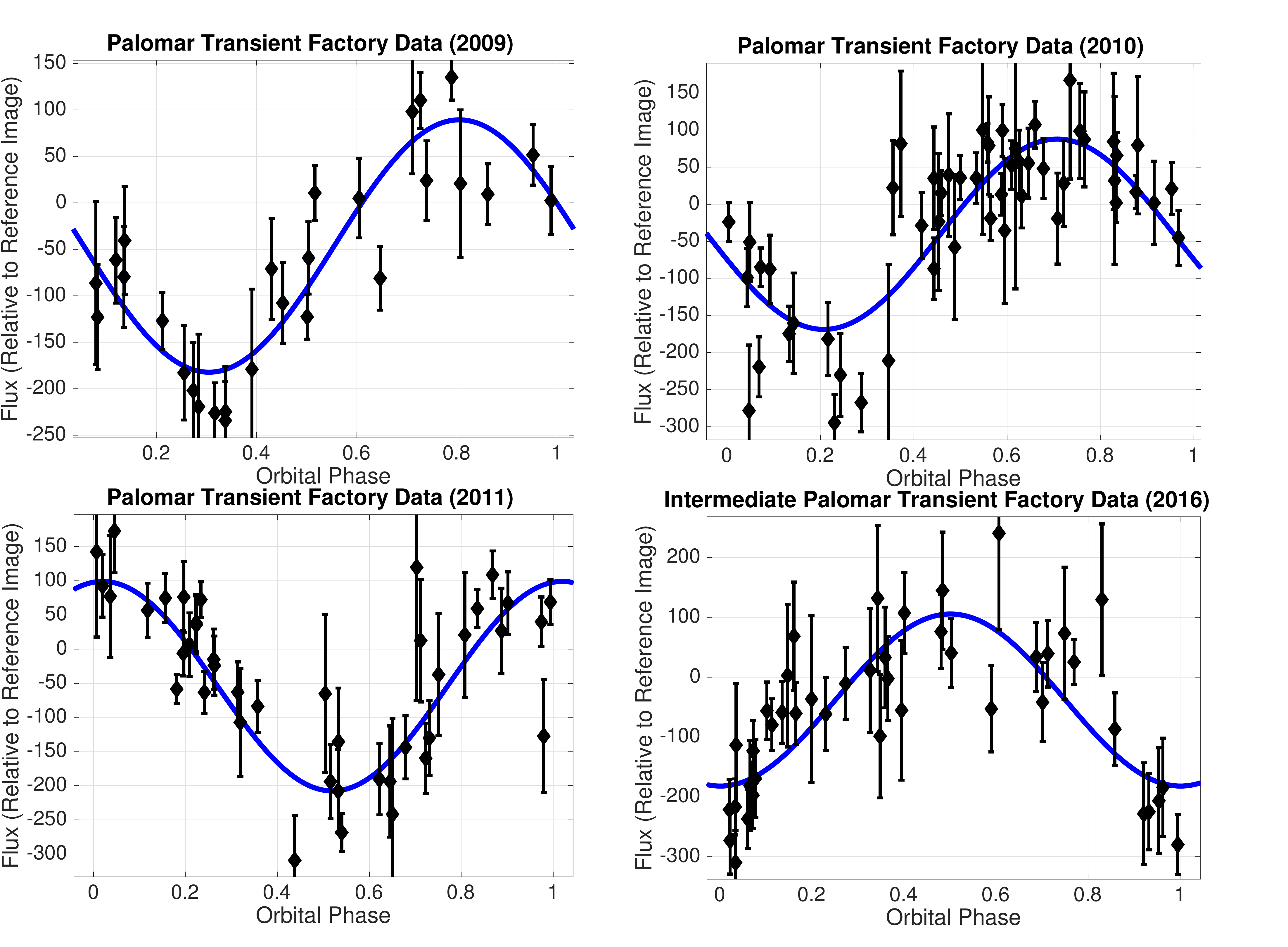}
\linespread{1.3}\selectfont{}
\caption{\textbf{Extended Data Figure 2: Fits to archival Palomar Transient Factory Data} Least squares fits of single harmonic sinusoids (smooth blue lines) to archival Palomar Transient Factory and Intermediate Palomar Transient Factory data used to determine the orbital decay rate. This archival data was extracted by using forced photometry on difference images. Because this is a least squares fit of a sinusoid to the data, this timing technique uses the reflection effect in the system as its primary clock, rather than the mid-eclipse time. All error bars are 1 sigma. To determine the time of the epoch, we take the mean of all epochs used, and then calculate the phase of eclipse nearest to this mean time.}
\label{fig:PTF}
\end{figure}

After obtaining the timing epochs, we measured the deviation of each eclipse time since the start of observations relative to a model with a constant orbital period (Figure 2). This was non-trivial, because the eclipse time had drifted by multiple orbits since the PTF epochs. However, upon allowing for an integer number of orbits to have passed, only one solution yielded a significant fit when using a quadratic model (corresponding to $5$ orbits since the initial epoch). A linear model failed in all cases, and a cubic model produced a cubic coefficient consistent with $0$, so for the final emphemeris, we chose to use a least squares fit of a quadratic to the these mid-eclipse times to estimate $\dot{P}$ (this fit is the red curve down in (Figure 3a), which resulted in an adjusted $R^2=0.999995$. This quadratic fit is independent of any model assumptions about decay due to tidal contribution, general relativity, or any other mechanism; it only indicates that the orbital period is decreasing with with an approximately constant rate as a function of time over the course of the observations.

In addition to extracting $\dot P$, we verified our measurement by fitting a quadratic to only mid-eclipse times extracted from CHIMERA and KPED data (Extended Data Figure 3), and obtained a value of $\dot P=(-2.487\pm0.19)\times10^{-11} \mathrm{s\,s}^{-1}$, consistent with the $\dot P=(-2.373\pm0.005)\times10^{-11} \mathrm{s\,s}^{-1}$ value obtained when including PTF/iPTF data. Moreover, these values are both consistent with the value of $\dot P=(-2.378\pm0.049)\times10^{-11} \mathrm{s\,s}^{-1}$ obtained by fitting only the four PTF/iPTF epochs and excluding all CHIMERA and KPED data.

\begin{figure}
\includegraphics[width=6.4in]{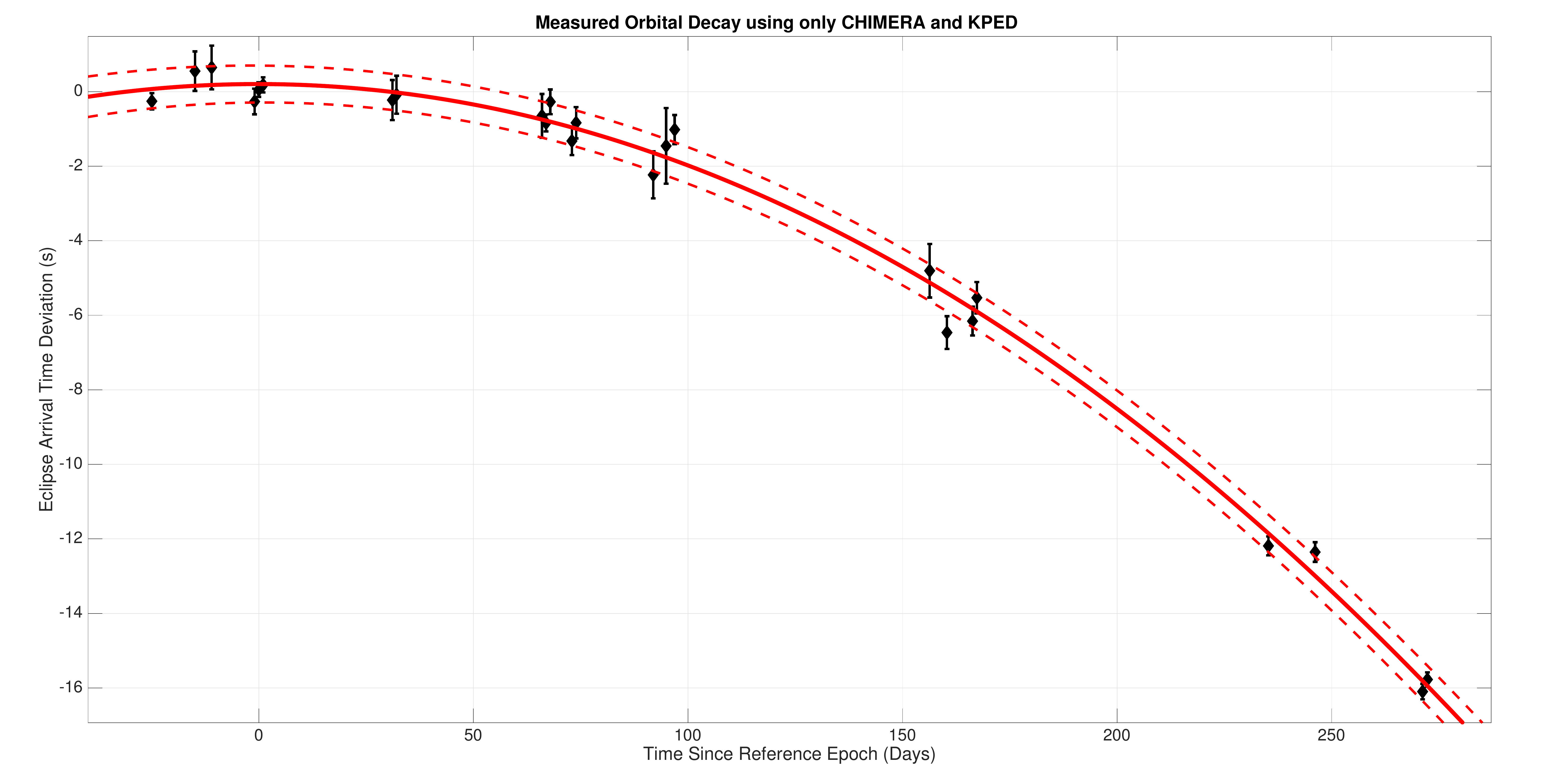}
\linespread{1.3}\selectfont{}
\caption{\textbf{Extended Data Figure 3: Orbital decay measured with CHIMERA and KPED} A quadratic fit (smooth red curve) to timing epochs exclusively originating from CHIMERA and KPED data (with the 68 percent confidence interval shown by the red dashed lines). This solution yielded a $\dot P$ consistent with the much more precise solution derived by including PTF/iPTF data. All error bars on the timing epochs are 1 sigma.}
\label{fig:CHIMERAKPED}
\end{figure}

With $\dot P$ measured, we assessed how much contribution there was due to ``secular acceleration"\cite{1970SvA....13..562S} and differential Galactic acceleration\cite{2012hpa..book.....L}.  For the former we used the proper motion as measured from the second data release of Gaia\cite{2018A&A...616A...1G}, $\mu=5.1\pm2.2\,{\rm mas}\,{\rm yr}^{-1}$.  This caused an apparent excess $\dot P_{\rm Shk}$ of $(8.3\pm3.6)\times 10^{-7} (d/1\,{\rm kpc})\,{\rm ms\,yr}^{-1}$ where $d$ is the distance. This term must be subtracted from the measured value to obtain the intrinsic $\dot P$.  Similarly, differential Galactic acceleration leads to an apparent $\dot P_{\rm DGR}$ of $-3.5\times 10^{-7}\,{\rm ms\,yr}^{-1}$  at a distance of $2.4$\,kpc, computed using the \texttt{MWPotential2014} potential\cite{2015ApJS..216...29B}.  Both of these contributions are significantly smaller than the measured $\dot P$ and are therefore negligible compared to the uncertainties from tidal contributions (Section 9).

\section{Spectroscopic Analysis}

To perform the spectroscopic analysis, first we coadded 317 individual spectra (all taken with a 52s exposure) into $12$ phase-bins, using the emphemeris for the mid-eclipse time of the primary eclipse to define phase $0$. We then fit stellar atmospheric models\cite{2017ApJS..231....1L} and obtained measurements of the logarithm of the surface gravity of the primary, $\log (g)_1$, and its effective photospheric temperature, $T_{\rm eff,1}$, using the spectrum immediately after the phase of the primary eclipse (phase $0.0833$), to minimize the flux contributed by the irradiated face of the secondary (Figure 3a).

In order to measure radial-velocity semi-amplitudes of the objects, $K_1$ and $K_2$, we used a double Gaussian model to fit the line associated with the hydrogen $n=5$ to $n=2$ transition in each phase-binned spectrum to extract the Doppler shifts of the emission and absorption lines.

To derive overall radial velocities, we adopted an out-of-phase sine wave model for the absorption and emission features
\begin{equation}
v_1(j) = K_{\rm 1measured} \sin \left(\frac{2 \pi j}{12} \right) + A, \hspace{1em}
v_2(j) = K_{\rm 2measured} \sin \left(\frac{2 \pi j}{12} + \pi \right) + B,
\label{eq:AB}
\end{equation}
where $j$ encodes the index of the phase-bin and $A$ and $B$ are the systemic velocities of each white dwarf, $K_{\rm 1measured}$ and $K_{\rm 2measured}$ are the observed velocity semi-amplitudes, and $v_1(j)$ and $v_2(j)$ are the Doppler shifted velocities associated with each phase-bin (Extended Data Figure 4). We required the two systemic velocities to satisfy $A>B$ when sampling, because the gravitational redshift of the primary should exceed that of the secondary because of the primary's larger surface gravity.
We used a Kernel Density Estimator (KDE) applied to the posterior distributions for the measured velocity semi-amplitudes $K_{\rm 1measured}$ and $K_{\rm 2measured}$ to assign the probabilities when sampling.

In order to derive masses from these spectroscopic fits, we applied corrections to the measured velocity semi-amplitudes. First, we applied a smearing correction of $20\%$ to both $K_{\rm 1measured}$ and $K_{\rm 2measured}$, due to our phase binned spectra each being co-additions of spectra taken over a third of the orbital phase (though each individual spectrum was taken over only an eight of the orbital phase, broader coadditions were necessary to reach sufficient SNR for measuring radial velocities). We used a prior of $C=0.68\pm0.04$ for the center of light correction term, based on the mass ratio estimate\cite{horne1989evidence}. This modified the velocity amplitude of the secondary by:

\begin{equation}
K_2=\frac{K_{\rm 2measured}}{1- C\frac{R_2}{a}(1+q)}
\label{eq:centeroflight}
\end{equation}
where $R_2/a$ was taken from the lightcurve fitting, and the mass ratio was defined as $q=K_1/K_2$ (thus requiring us to solve this expression to isolate $K_2$). We applied the center-of-light correction to the secondary, because one face of it is heavily irradiated (meaning that the emission lines arise from a location different than the center of mass).

\begin{figure}
    \centering
    \includegraphics[width=6.2in]{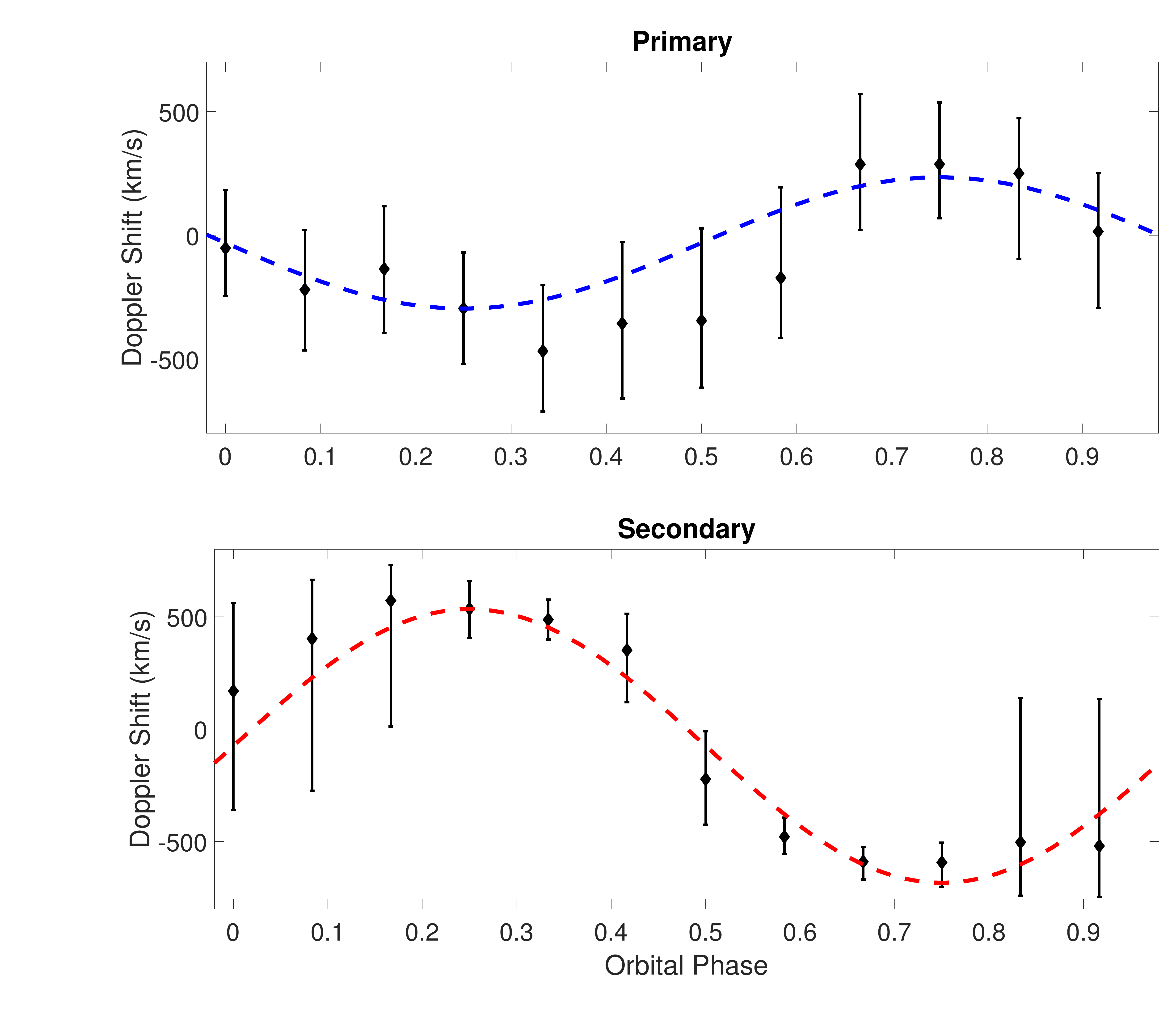}
    \linespread{1.3}\selectfont{}
    \caption{\textbf{Extended Data Figure 4: Radial velocity measurements of ZTFJ1539+5027} A plot of the measured Doppler shifts vs. orbital phase for the primary and secondary. The primary eclipse occurs at orbital phase $0$. In the top panel, we plot measured Doppler shifts of the more massive primary, extracted from 12 phase bins of coadded spectra. The dashed blue line illustrates the fit of a sinusoid to this data (adjusted $R^2=0.7118$). The lower panel shows the Doppler shift measurements of the secondary, and also the best fit sinusoid to this data (adjusted $R^2=0.9757$). Because of the low SNR of the spectra, these fits have large uncertainties (especially in the case of the primary, with its shallow and broad absorption lines). This is reflected in the broad distribution of possible masses associated with the spectroscopic constraint illustrated in Figure \ref{fig:combined}. All error bars are 68\% CIs.}
    \label{fig:RV}
\end{figure}

\section{Mass and Radius Analysis}

There are three main constraints that contribute to the overall mass and radius estimates.

\subsection{6.1   Spectroscopic Constraint:} Using Kepler's law and Classical Mechanics, we can relate the velocity semi-amplitude, inclination, and period of a binary to the masses with a binary mass function:

\begin{equation}
    \label{ddot}
    \frac{M_2^3{\sin}^3(i)}{(M_1+M_2)^2}=\frac{PK_1^3}{2\pi G}
\end{equation}where $G$ is the gravitational constant. Because we have measured both $K_1$ and $K_2$, as well as $P$ and $i$, we can write two such equations, and derive constraints on the two masses  (Figure 4).
\subsection{6.2   White Dwarf Model Constraint:} We used a mass-radius relation from models for a carbon-oxygen white dwarf\cite{AlCa2015}, which depends on mass, metallicity, hydrogen fractional mass in the atmosphere, and radius, together with the spectroscopic measurements of $T_{\rm eff,1}$ and $\log (g)_1$, to derive constraints on these properties. In these fits, we marginalized over metalicity and hydrogen mass fraction. We used the measured ratio of the radius of the primary to the semi-major axis $\frac{R_1}{a}$ (from the lightcurve modelling), and combined this with Kepler's law, and the white dwarf model constraints\cite{AlCa2015}, to constrain the system parameters. This also weakly constrains the mass of the secondary, which enters into Kepler's law as contributing to the total mass of the system.
\subsection{6.3   Chirp Mass Fit:} We used the measured $\dot{f}$ to infer the chirp mass, $M_c=\frac{(M_1M_2)^{\frac{3}{5}}}{(M_1+M_2)^{\frac{1}{5}}}$, which is be related to the orbital decay rate by Equation 5: 

\begin{equation}
   \dot{f}=\frac{96}{5}\pi^{\frac{8}{3}} \left(\frac{GM_c}{c^3}\right)^\frac{5}{3}f^\frac{11}{3}
\end{equation}where $c$ is the speed of light. Because the chirp mass inferred from $\dot{f}$ assumes the decay is caused purely by general relativity, we also computed a chirp mass assuming a 10 percent tidal contribution to $\dot{f}$ as a lower bound on $M_c$.

\subsection{6.4   Combined Fit:}To combine all of the constraints described above, we created a KDE based on the $M_1$ and $M_2$ estimates from the spectroscopic, white dwarf model, and chirp mass constraints. The posterior distribution of this combined analysis yielded the values for the masses reported in Table 1. The radii were computed using $r_1$ and $r_2$ from the lightcurve modelling, and the semi-major axis $a$ determined with the masses and the orbital period.

\section{Distance Estimate}

Because no reliable parallax measurement exists for ZTF J1539+5027, we instead used its bolometric luminosity to estimate the distance. First, we measured the apparent magnitude of the hot primary without contribution from the irradiated face of the secondary by performing a least squares fit of a sinusoid to the ZTF-$g$ lightcurve, omitting data from the eclipse, and measuring the minimum of this sinusoid. This yields an apparent ZTF-$g$ band apparent magnitude of $20.38\pm0.05$.

Next, we used the atmospherically determined temperature of the primary, $T_{\rm eff,1}\approx 48,900 \, {\rm K}$, to infer the absolute $g^\prime$ luminosity of the $0.6$\,\(M_\odot\) DA white dwarf\cite{holberg2006calibration}. The spectroscopic temperature puts ZTF J1539+5027 approximately between two temperatures in the models--one corresponding to $T_{\rm eff,1}=45,000\, {\rm K}$ and the other to $T_{\rm eff,1}=55,000\, {\rm K}$. These corresponded to $g^\prime$ absolute magnitudes of $8.71$ and $8.35$, respectively, and we adopted these values as the lower and upper bounds on the object's absolute luminosity, and assume a uniform distribution of possible absolute luminosities between these values. Using this to solve for the distance of the object, we estimated a distance of $d=2.34\pm0.14\,\rm{kpc}$, though we emphasize that the error bars derived using atmospherically determined quantities tend to be underestimated. We also incorporated a uniform distribution of $E(B-V)$ in the range of $0$ to $0.04$ to account for the effects of extinction at these coordinates\cite{green2018galactic}.

\section{Gravitational Wave Strain}

The expression for the characteristic strain\cite{korol2017prospects} used in Figure 2b (including the value plotted for ZTF J1539+5027) is:
\begin{equation}
   S_{c}=\frac{2(GM_{c})^{5/3}(\pi f)^{2/3}}{c^{4}d}\sqrt{fT_{obs}}
\end{equation}
where $d$ is the distance to the object, $T_{obs}$ is the integrated observation time of the LISA mission. Though it is the conventional quantity used to construct such diagrams, the characteristic strain does not capture any information about source inclination, detector response, etc. 

In order to compute the signal to noise\cite{korol2017prospects} for LISA, we directly invoked the signal amplitude at the detector $A=\sqrt{|F_{+}|^{2}|h_{+}|^{2}+|F_{\times}|^{2}|h_{\times}|^{2}}$, where $h_{+}$ and $h_{\times}$ are the two gravitational wave polarization amplitudes, and $F_{+}$ and $F_{\times}$ are the LISA detector response patterns corresponding to these polarizations. The $h_{+}$ polarization amplitude includes a factor of $(1+\cos^2{(i)})$ and the $h_{\times}$ polarization a factor of $2\cos{(i)}$, meaning that systems like ZTF J1539+5027 with an inclination close to 90 degrees exhibit a gravitational wave signal up to a factor of $\sqrt{8}$ smaller than an equivalent face-on system with an inclination close to 0 degrees in situations where $F_{+}\approx F_{\times}$. In Table 1, we include an estimate for both the SNR of ZTF J1539+5027, and SDSS J0651+2844 computed using the same technique.

\section{Tidal Effects}

\subsection{Tidal Contribution to Orbital Decay}

Tidal dissipation can transfer orbital energy into rotational and thermal energy within the stars. The former effect can cause the orbit to decay slightly faster than gravitational radiation alone, while the latter effect can increase the surface temperatures of the stars. Studies of tidal synchronization and heating predict that tidal energy dissipation scales more strongly with orbital period than gravitational radiation\cite{fullerwd:12,fuller2013dynamical}. As the orbit decays and the white dwarfs draw nearer to each other, tides begin to act on a shorter timescale than orbital decay, spinning up the stars toward synchronous rotation. The ``critical" orbital period, $P_c$, below which tidal spin-up can occur faster than orbital decay, lies in the range $P_c \sim 45-130$ minutes\cite{fullerwd:12}, depending on the white dwarf mass and age. In any case, at an orbital period of only $6.91$ minutes, we expect the stars in ZTF J1539+5027 to be spinning synchronously with the orbit. In this regime, the rate of tidal energy transfer from the orbit to the stellar interiors is 
\begin{equation}
    \label{edot}
    \dot{E}_{\rm tide} \simeq 4 \pi^2 I \frac{\dot{P}_{\rm GW}}{P^3} \, ,
\end{equation}
where $I = I_1 + I_2$ is the combined moment of inertia of the two stars, and $\dot{P}_{\rm GW}$ is the orbital period decay caused by gravitational waves.

Comparing Equation \ref{edot} with the orbital energy lost to gravitational wave emission, the tidal contribution $\dot{P}_{\rm tide}$ to the total orbital decay rate $\dot{P}$ is given by
\begin{equation}
    \label{pdottide}
    \frac{\dot{P}_{\rm tide}}{\dot{P}_{\rm GW}} \simeq \frac{ 12 \pi^2 I a}{G M_1 M_2 P^2} \, .
\end{equation}
The moment of inertia of star 1 is $I_1 = \kappa_1 M_1 R_1^2$ (and similarly for star 2), where $\kappa_1$ is a dimensionless constant determined by the internal structure of the white dwarf. Equation \ref{pdottide} can also be written as
\begin{equation}
    \label{pdottide2}
    \frac{\dot{P}_{\rm tide}}{\dot{P}_{\rm GW}} \simeq \frac{ 3 (M_{1}+M_{2})^2}{M_1 M_2} \bigg[ \kappa_1 \frac{M_1}{M_1 + M_2} \bigg( \frac{R_1}{a}\bigg)^2 + \kappa_2 \frac{M_2}{M_1 + M_2} \bigg( \frac{R_2}{a}\bigg)^2 \bigg] \, .
\end{equation}
From the white dwarf models (Section \ref{binary}), we find $\kappa_1 \simeq 0.14$ and $\kappa_2 \simeq 0.11$. Using  $M_1 \simeq 0.61 M_\odot$, $M_2 \simeq 0.21 M_\odot$, $R_1/a \simeq 0.14$, $R_2/a \simeq 0.28$, we find $\dot{P}_{\rm tide}/\dot{P}_{\rm GW} \simeq 0.067$. Thus, we expect the orbit to decay several percent faster than gravitational radiation alone, provided that the stars are tidally locked. There is an additional effect of orbital decay caused by orbital energy used to raise a tidal bulge in each star \cite{benacquista:11,shah:13}, but we find it to be more than an order of magnitude smaller than Equation \ref{pdottide2}.

\subsection{Tidal Heating}

The tidal energy dissipation within the white dwarfs is partitioned between kinetic energy used to spin up the white dwarfs and heat that can diffuse out and increase their surface temperatures. For an aligned, sub-synchronous and rigidly rotating white dwarf, the ratio of tidal heating to tidal energy dissipation is  $\dot{E}_{\rm heat}/\dot{E}_{\rm tide} = 1 - P/P_{\rm spin}$, where $P_{\rm spin}$ is the spin period of the white dwarf\cite{fullerwd:12}. At orbital periods below the critical value $P_c$, the spin period, $P_{\rm spin}$, decreases to approach the orbital period. Calculations suggest that the tidal heating rate in this regime is expected to be $\dot{E}_{\rm heat} \sim \dot{E}_{\rm tide} (P/P_c)$\cite{fuller2013dynamical}, such that the tidal heating rate is much smaller than the energy dissipation rate, because most of the energy is converted to rotational energy. 

To estimate an upper limit on the surface temperature of each white dwarf that can be obtained from tidal heating alone, we assume $\dot{E}_{\rm heat} \sim \dot{E}_{\rm tide} \approx 6 \pi^2 I /(P^2 t_{\rm GW})$, where $t_{\rm GW} = \frac{3}{2}\frac{P}{|\dot{P}|}$ is the gravitational wave timescale. If the tidal heat is instantaneously reradiated, this corresponds to a surface temperature
\begin{align}
    \label{ttide}
    T_{\rm tide} &= \bigg( \frac{\dot{E}_{\rm heat}}{4 \pi \sigma_B R^2} \bigg)^{1/4} \nonumber \\
    &= \bigg( \frac{3 \pi \kappa M }{2 \sigma_B P^2 t_{\rm GW}} \bigg)^{1/4} \, ,
\end{align}
where $\sigma_B$ is the Stefan-Boltzmann constant, and $M$ is the mass of the white dwarf. Note that this temperature depends only weakly on the white dwarf mass and moment of inertia, and is independent of the white dwarf radius. Using the same values as above and the measured value $t_{\rm GW}  \approx 830,000 \, {\rm yr}$, we find an upper limit of $T_{\rm tide} \approx 44,000 \, {\rm K}$ for the primary white dwarf. However, accounting for the suppression factor $P/P_c \sim 6.9/60$, such that $\dot{E}_{\rm heat} \sim \dot{E}_{\rm tide} (P/P_c)$, yields a more realistic temperature $T_{\rm tide} \approx 25,000 \, {\rm K}$. Hence, while tidal heating may be able to heat the primary to temperatures near that observed, our best estimate suggests substantially lower temperatures. Applying Equation \ref{ttide} to the secondary white dwarf predicts an upper limit due to tidal heating of $T_{\rm tide} \approx 33,000 \, {\rm K}$, but our best estimate is $T_{\rm tide} \approx 19,000 \, {\rm K}$. A measured nightside temperature near this value would be consistent with tidal heat powering the nightside flux of the secondary. However, the value of the surface brightness ratio in $g^\prime$ suggests a secondary temperature of $T_{\rm eff,2}<10,000\, {\rm K}$.

\section{Binary Models}
\label{binary}

To understand the evolutionary state of ZTF J1539+5027, we have constructed several binary models using the MESA stellar evolution code\cite{paxton:15}. The models contain a helium white dwarf of $0.2$ or $0.25 \, M_\odot$ with a point mass companion of $0.6$ or $0.55 \, M_\odot$. The mass of the hydrogen envelope of the helium white dwarf is then reduced to the values shown in Extended Data Figure 5. We initialize these runs at orbital periods of $1$ hour to mimic the end of a common envelope event that formed the tight binary. The helium white dwarf is evolved simultaneously with the orbit, with angular momentum losses due to gravitational waves and fully non-conservative mass transfer. 

\begin{figure}
    \centering
    \includegraphics[width=5in]{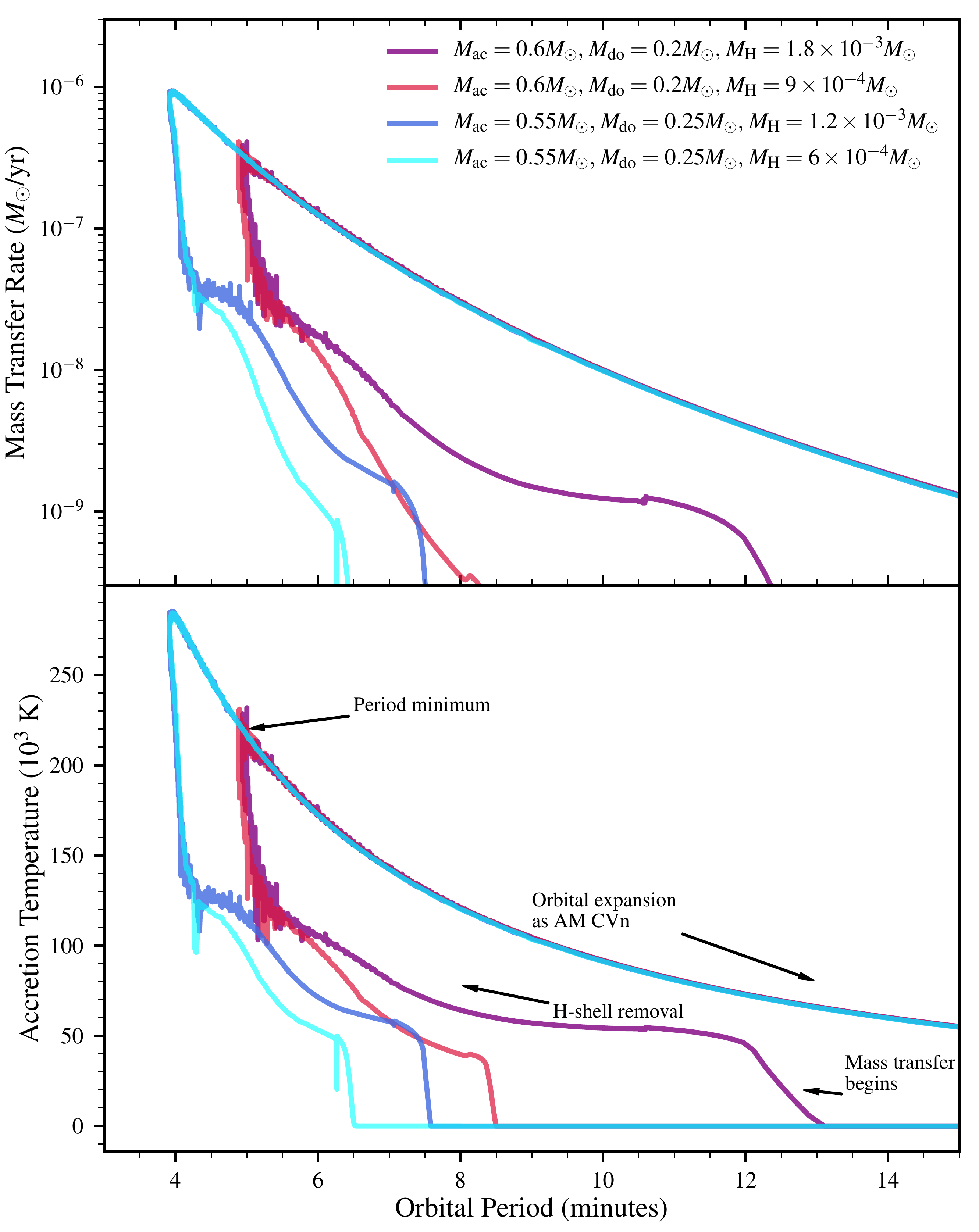}
    \linespread{1.3}\selectfont{}
    \caption{\textbf{Extended Data Figure 5: Binary evolution models} Binary stellar evolution models for systems similar to ZTF J1539+5027. {\bf Top:} Mass transfer rate as a function of orbital period. Systems begin at large orbital period and move towards smaller period due to gravitational radiation, and in some cases they move back out due to stable mass transfer. Except for high-mass donors with thin hydrogen envelopes, mass transfer is expected to begin at orbital periods longer than 7 minutes. {\bf Bottom:} Corresponding accretion temperature from Equation \ref{Tac}. }
    \label{fig:models}
\end{figure}

Extended Data Figure 5 shows a plot of the mass loss rate from the secondary as function of orbital period. The secondary overflows its Roche lobe and begins mass transfer at orbital periods ranging from $P\sim 6.5-13$ minutes, and the expected mass loss rates at $P= 7$ minutes are typically $\dot{M} \sim 3 \times 10^{-9} \, M _\odot \, {\rm yr}^{-1}$ (ranging from $0$ up to $10^{-8} \, M _\odot \, {\rm yr}^{-1}$, depending on the white dwarf masses and hydrogen envelope mass). During the initial phase of slow mass transfer, the secondary loses its non-degenerate hydrogen envelope as the Roche lobe contracts inward. The mass accreting onto the primary can greatly heat it as gravitational potential energy is converted to heat. The energy released by accretion is $\dot{E}_{\rm accrete} \sim G M_1 \dot{M}/R_1$, with order unity corrections due to its non-zero kinetic and gravitational energy when it is lost from the secondary. We do not evolve the primary (accretor), but we may crudely estimate its temperature by assuming that the accretion energy is uniformly radiated as a blackbody over its surface. The corresponding accretion temperature (assuming 100\% efficiency) is
\begin{equation}
    \label{Tac}
    T_{\rm accretion} = \bigg( \frac{G M_1 \dot{M}}{4 \pi \sigma_B R^3} \bigg)^{1/4} \, .
\end{equation}
The bottom panel of Extended Data Figure 5 shows that once mass transfer begins, it can easily increase the primary's temperature to $T_1 \gtrsim 50,000\, {\rm K}$.

\section{Constraint on Accretion}

Although accretion could explain the high temperature of the primary, we have not detected any evidence of ongoing mass-transfer. We have constrained the possibility of active accretion contributing luminosity to an accretion hot spot using both optical and X-ray data. The upper limit we inferred is $\dot{M}< 2\times10^{-8} M_\odot \,{\rm yr}^{-1}$, and this is only possible in a very narrow hot spot temperature range (Extended Data Figure 6). For both the optical and X-ray constraint we have assumed that only $10\%$ of accretion energy is being converted to the luminosity of the hot spot. This is based on a model where we assume $90\%$ of the accretion energy is being deposited into heating the optically thick photosphere of the white dwarf, while only $10\%$ is contributing to luminosity immediatly re-radiated in the form of a hot spot. We estimated the upper limit by computing the X-ray flux using NASA's WebPIMMS mission count rate simulator, using a 3-sigma upper limit on the background count rate determined from the X-ray images. We assumed a hydrogen column density\cite{kalberla2005leiden} of $\rm nH=1.5\times 10^{20}\,\rm cm^{-2}$. We used the distance of $d=2.4\, \rm kpc$ to convert the upper limit on the unabsorbed X-ray flux to an X-ray luminosity. For the optical constraint, we computed an upper limit on the optical luminosity of the hot spot as $10\%$ the luminosity of the photosphere of the hot primary, based on an absence of emission lines in our coadded spectra, which have an SNR of approximately 10. We calculated the upper limit on $\dot{M}$ at various temperatures by integrating a Planck function at the corresponding temperature over the instrument passbands, and then computed the maximum bolometric luminosity, $L_{\rm accretion}$, an emitting region at this temperature could have and still be consistent with the non-detections in these passbands. We then determined the corresponding upper limit on $\dot{M}$ by equating $L_{\rm accretion}=0.1GM_1\dot{M}/R_1$. We obtained the XMM EPIC-pn data from the public XMM-Newton science archive (observation ID: 0800971501 PI: Pratt, Gabriel). The SWIFT observation was obtained with our own program (observation ID: 00010787001, 	00010787002 PI: Kulkarni).

\begin{figure}
    \centering
    \includegraphics[width=5.4in]{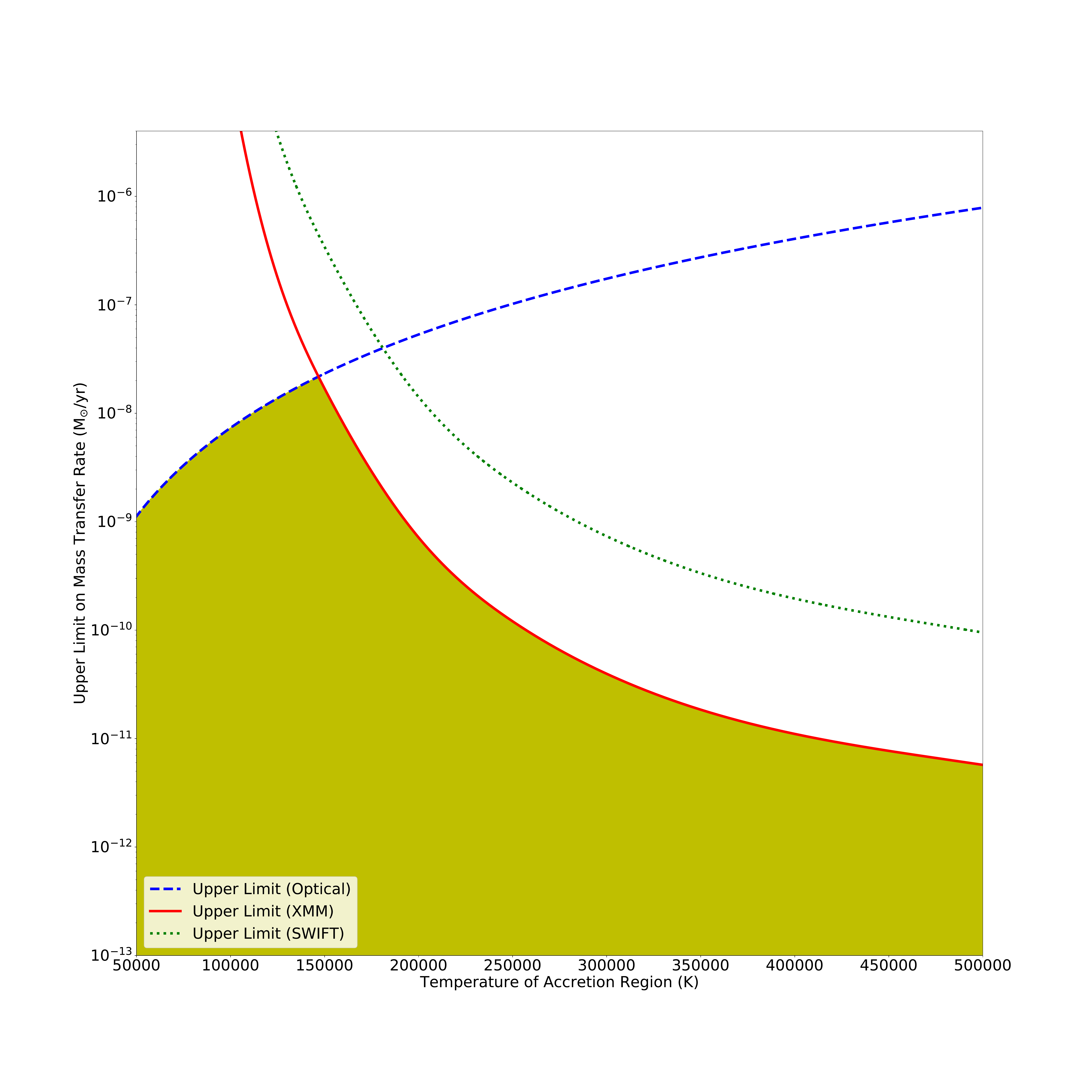}
    \linespread{1.3}\selectfont{}
    \caption{\textbf{Extended Data Figure 6: X-ray and optical constraints on accretion in ZTFJ1539+5027} The constraints on mass transfer resulting from the non-detection of any signatures of accretion in both the optical and X-ray bands. The upper limits are expressed in terms of the mass accretion rate contributing to the accretion luminosity of a hypothetical hot spot. The solid red curve illustrates the constraint imposed by the XMM EPIC-pn X-ray non-detection, which rules out significant mass transfer contributing to a hot spot with temperatures greater than $\approx 150,000\, {\rm K}$, while the green dotted line illustrates a weaker upper limit imposed by the non-detection in a SWIFT XRT observation. We constructed the dashed blue curve, which represents the optical constraint, by requiring that any accretion luminosity originating from a hot spot should contribute $<10\%$ to the luminosity in the band ranging from $320$ to $540\, {\rm nm}$, as we know from the optical spectrum (Figure \ref{fig:spectra}) that this light is dominated by the $\approx 50,000\, {\rm K}$ photosphere of the hot primary, and also see no signature of a hot spot in the CHIMERA lightcurve (Figure \ref{fig:chimera}). We chose the threshold of $<10\%$, because given the SNR of the spectra, we expect we should be able to detect optically thin emission with an amplitude at the $10\%$ level. Other white dwarfs with such a hot spot (such as HM Cancri) exhibit such emission, particularly in lines associated with ionized helium. }
    \label{fig:accretion}
\end{figure}

\section{Novae}

The high temperature of the primary may plausibly be explained if the white dwarf is cooling after having recently undergone a nova outburst, caused by accretion of hydrogen from the secondary or a tidally induced nova \cite{fuller2012tidal}. The nova likely ejects an amount of mass comparable to the hydrogen shell mass which must be accreted, which for a $0.6 M_\odot$ white dwarf is $M_H \sim 10^{-4} \, M_\odot$ \cite{wolf:13}. Following the nova, the orbit widens slightly, and the system is brought out of contact such that mass transfer from the secondary ceases. The change in semi-major axis following the loss of the nova shell is $\Delta a/a \sim M_H/(M_1+M_2)$.
The length of time the binary is detached before gravitational wave emission brings the system back into contact is
\begin{equation}
    t_{\rm detach} = t_{\rm GW} \frac{\Delta a}{a} \sim t_{\rm GW} \frac{M_H}{M_1 + M_2} \, .
\end{equation}
Using $M_H =10^{-4} \, M_\odot$, we estimate $t_{\rm detach} \sim 100$ yr. This can be compared to the time spent accreting mass between subsequent novae. 
\begin{equation}
    t_{\rm accrete} = \frac{M_H}{\dot{M}} \, .
\end{equation}
As the non-degenerate hydrogen envelope of the low-mass secondary is stripped off (see discussion in\cite{kaplan:12}), the approximate mass transfer rate is expected to be $\dot{M} \lesssim 10^{-8} M_\odot \, {\rm yr}^{-1}$ (Extended Data Figure 5). The time between novae outbursts is thus $t_{\rm accrete} \gtrsim 10^{4}$ years. Then the ratio of time spent in a detached state relative to an accreting state is $t_{\rm detach}/t_{\rm accrete} < 10^{-2}$. Hence, while it is possible that the system is in a detached state following a nova caused by mass transfer, the chances of catching the system in this state are small. To help rule out the possibility, we used the WASP instrument on the Hale telescope to obtain a deep H-$\alpha$ image of the field and found no evidence for a remnant nova shell; however, this analysis was limited by the lack of an off-band image.

\section{Population Implications}

From \cite{brown:16}, the merger rate of He+CO WDs in the Milky Way is roughly $0.003 \, {\rm yr}^{-1}$. This number is reached from both observational and population synthesis arguments. The number of systems with decay time equal or less than the $\sim$210 kyr decay time of ZTF J1539+5027 is thus $\sim$630. Out to the distance of 2.3 kpc, given a local surface density of $68 \, M_\odot \, {\rm pc}^{-2}$ from \cite{bovy:13}, the stellar mass is $\sim \! 10^9 \, M_\odot$, roughly $2\%$ of the total disk mass of $\sim  \! 5 \times 10^9 \, M_\odot$. We thus expect to find $\sim$13 binaries with a similar distance and merging timescale as ZTF J1539+5027. The fraction of eclipsing systems is roughly $R/a \sim 0.25$ for our measured parameters, hence we may expect $\sim$3 eclipsing systems like ZTF J1539+5027. ZTF can detect such systems in most of the volume out to its distance, as long as they are as bright as this system. We may be missing slightly longer period systems that are dimmer because they have not yet started mass transfer. We comment that the estimate from \cite{brown:16} found that many double WDs must be born at short orbital periods in order to explain the abundance of short period systems relative to longer period systems, and ZTF J1539+5027 may support that conclusion. 

\begin{figure}
    \centering
    \includegraphics[width=7in]{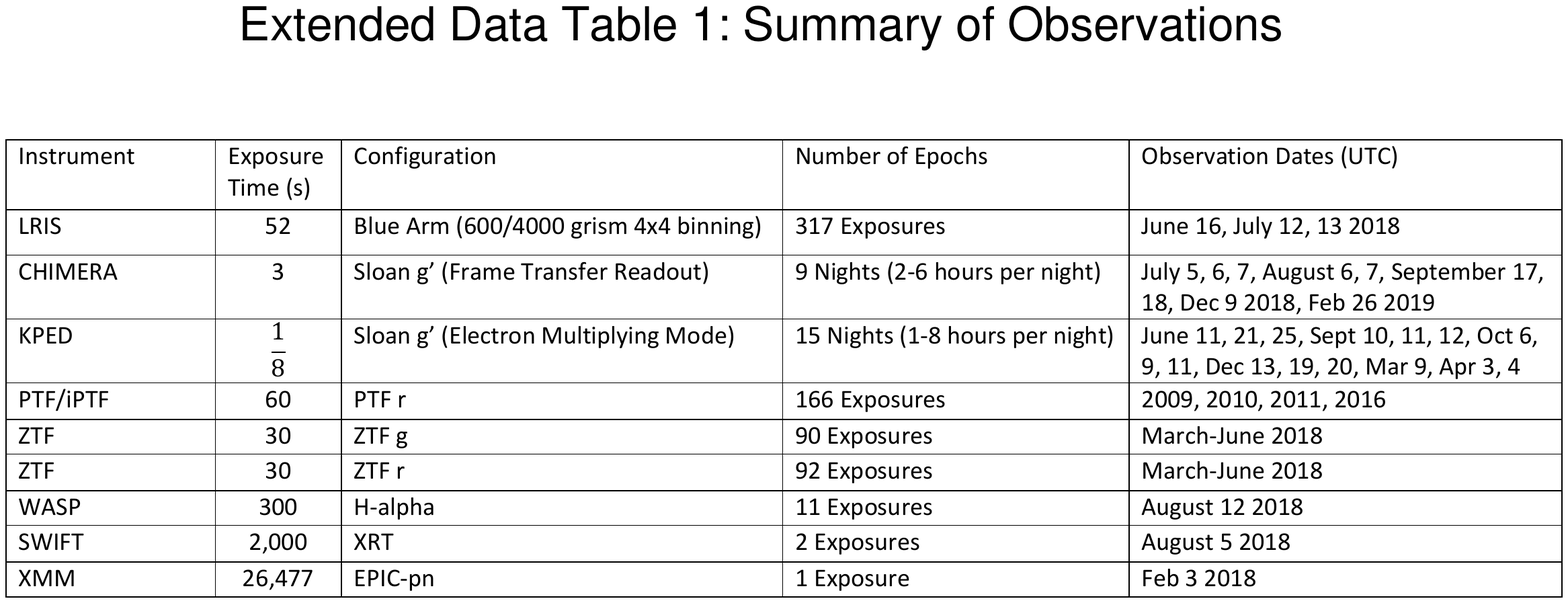}
    \label{fig:table2}
\end{figure}

\end{methods}

\section{Data Availability}
Upon request, the first author will provide reduced photometric and spectroscopic data, and available ZTF data for the object. We have included the eclipse time data used to construct the orbital decay diagram in Figures 2a, Extended Data Figure 2, and Extended Data Figure 3. The X-ray observations are already in the public domain, and their observation IDs have been supplied in the text. The proprietary period for the spectroscopic data will expire at the start of 2020, at which point this data will also be public and readily accessible. 

\section{Code Availability}

Upon request, the first author will provide code (primarily in python) used to analyze the observations and data such as posterior distributions used to produce the figures in the text (MATLAB was used to generate most of the figures). 

\begin{addendum}
 \item KBB thanks the National Aeronautics and Space Administration and the Heising Simons Foundation for supporting his research. 
 
Based on observations obtained with the Samuel Oschin Telescope 48-inch and the 60-inch Telescope at the Palomar Observatory as part of the Zwicky Transient Facility project. ZTF is supported by the National Science Foundation under Grant No. AST-1440341 and a collaboration including Caltech, IPAC, the Weizmann Institute for Science, the Oskar Klein Center at Stockholm University, the University of Maryland, the University of Washington (UW), Deutsches Elektronen-Synchrotron and Humboldt University, Los Alamos National Laboratories, the TANGO Consortium of Taiwan, the University of Wisconsin at Milwaukee, and Lawrence Berkeley National Laboratories. Operations are conducted by Caltech Optical Observatories, IPAC, and UW.
 
The KPED team thanks the National Science Foundation and the National Optical Astronomical Observatory for making the Kitt Peak 2.1-m telescope available. The KPED team thanks the National Science Foundation, the National Optical Astronomical Observatory and the Murty family for support in the building and operation of KPED. In addition, they thank the CHIMERA project for use of the Electron Multiplying CCD (EMCCD).

 Some of the data presented herein were obtained at the W.M. Keck Observatory, which is operated as a scientific partnership among the California Institute of Technology, the University of California and the National Aeronautics and Space Administration. The Observatory was made possible by the generous financial support of the W.M. Keck Foundation. The authors wish to recognize and acknowledge the very significant cultural role and reverence that the summit of Mauna Kea has always had within the indigenous Hawaiian community. We are most fortunate to have the opportunity to conduct observations from this mountain.
 
This research benefited from interactions at the ZTF Theory Network Meeting that were funded 
by the Gordon and Betty Moore Foundation through Grant GBMF5076 and support from the National Science Foundation
through PHY-1748958 
 
 We thank John Hoffman, the creator of \texttt{cuvarbase}. We thank Thomas Marsh, Sterl Phinney, and Valeryia Korol for valuable discussions. We thank Gregg Hallinan and Christoffer Fremling for helping observe the object.

 \item[Competing Interests] The authors declare that they have no
competing financial interests.
\item[Contributions] KBB discovered the object, conducted the lightcurve analysis, eclipse time analysis, and was the primary author of the manuscript. KBB and MWC conducted the spectroscopic analysis. KBB, MWC, and TAP conducted the combined mass-radius analysis. KBB and MWC reduced the optical data. KBB, MWC, and DLK reduced and analysed the X-ray observations. JF conducted the theoretical analysis, including that on tides, and MESA  evolutionary models. KBB, MWC, TK, SRK, JvR, and TAP all contributed to collecting data on the object. KBB, MWC, JF, TK, ECB, LB, MJG, DLK, JvR, SRK, and TAP contributed to the physical interpretation of the object. TK, ECB, RGD, MF, MG, SK, RRL, AAM, FJM, RR, DLS, MTS, RMS, PS and RW contributed to the implementation of ZTF; MJG is the project scientist, TAP and GH are Co-PIs, and SRK is PI of ZTF. RGD, DAD, MF, RR contributed to the implementation of KPED; MWC is project scientist, and SRK is PI of KPED. TAP is KBB's PhD advisor.
 \item[Correspondence] Correspondence and requests for materials
should be addressed to  Kevin B Burdge~(email: kburdge@caltech.edu)
\end{addendum}

\end{document}